\documentclass[nofootinbib,physrev,twocolumn]{revtex4-2}

\pdfoutput=1
\usepackage{graphicx}
\usepackage{dcolumn}
\usepackage{bm}
\usepackage{amsmath}
\usepackage{hyperref}
\usepackage{mathtools}
\usepackage{amssymb}
\usepackage{physics}
\usepackage{slashed}
\usepackage{simpler-wick}
\usepackage{cancel}
\usepackage{xcolor}
\usepackage{bibunits}

\defaultbibliographystyle{apsrev4-2}
\defaultbibliography{ajarsystems}  

\begin{document}
\preprint{APS/123-QED}

\title{\textbf{Effective description of ajar systems with a \texorpdfstring{$U(1)$}{U(1)} symmetry} 
}%

\author{Can Onur Akyuz}
\email{cakyuz@andrew.cmu.edu}
\author{Riccardo Penco}
\email{rpenco@andrew.cmu.edu }
\affiliation{%
Department of Physics, Carnegie Mellon University, Pittsburgh, Pennsylvania 15213, USA}%

\date{\today}

\begin{abstract}
 We introduce the concept of ``ajar systems'' as an intermediate case between closed and open systems, where the time scale for charge exchange with the environment is parametrically larger than all other characteristic time scales. The Schwinger-Keldysh effective action for such finite-temperature systems exhibits a symmetry group $G_1 \times G_2$ weakly broken to its diagonal subgroup $G_{\text{diag}}$, which we systematically describe using spurion techniques. For $G = U(1)$, we calculate leading-order corrections to correlation functions in both diffusive and spontaneously broken phases. Unlike systems with approximate symmetries where both real and imaginary parts of dispersion relations receive corrections, ajar systems modify only the imaginary part---preserving gapless modes while adding finite damping.
\end{abstract}

\maketitle

\begin{bibunit}
\section{Introduction}

The application of effective field theory techniques to hydrodynamic systems and, more in general, systems out of equilibrium has recently received significant attention (see e.g.~\cite{Haehl:2016pec,Liu:2018kfw,Salcedo:2024smn}). The relevant framework to discuss these issues is the Schwinger-Keldysh (or ``in-in'') formalism~\cite{kamenev2023} and generalizations thereof~\cite{Haehl:2024pqu}.

Previous work~\cite{Sieberer:2015svu, Akyuz:2023lsm} showed that a local Schwinger-Keldysh effective action at low energies ($E \ll T$) for a \emph{closed} system at finite temperature $T$ must possess two copies of all global internal symmetries, forming a group $G_1 \times G_2$, partially realized nonlinearly. The individual symmetries $G_1$, $G_2$ do not have a direct physical interpretation, however the Noether current of the off-diagonal symmetry  corresponds to the physically observed currents, and their (non)conservation can be used to diagnose the openness of the system~\cite{Sieberer:2015svu}.  More precisely, an \emph{open} system, which exchanges conserved charges with an environment, has an effective action invariant only under the diagonal subgroup $G_{\rm diag}$~\cite{Sieberer:2015svu, Hongo:2019qhi}, reflecting the fact that charge conservation is now only satisfied on average in equilibrium, and not as an exact operator statement.

This distinction suggests an intermediate scenario, termed \emph{ajar} systems, where interactions with the environment occur over time scales much longer than internal dynamics. Such systems experience weak explicit symmetry breaking from $G_1 \times G_2$ to $G_{\rm diag}$, that can be  systematically modeled via a spurion technique~\cite{Penco:2020kvy}. Introducing a spurion field transforming bilinearly under $G_1 \times G_2$,
\begin{align}
O \rightarrow U_1 O U_2^{\dag }\, ,
\end{align}
and assigning it an expectation value $\langle O \rangle = \gamma$ invariant under $G_{\rm diag}$ allows systematic accounting of symmetry-breaking effects. The parameter $\gamma$ characterizes the interaction strength with the environment; small $\gamma$ relative to the EFT cutoff ensures approximate symmetry, yielding predictions for corrections to observables.

Before proceeding, we emphasize the difference between our approach and that of~\cite{Hongo:2024brb}. The latter studied the hydrodynamic behavior of closed systems exhibiting approximate symmetries (see also~\cite{Baggioli:2023tlc}), introducing two spurion fields $O_1$ and $O_2$ transforming as
\begin{align}
	 O_1 \to U_1 O_1, \quad O_2 \to U_2 O_2.
\end{align}
When these fields acquire nonzero expectation values, the full symmetry group $G_1 \times G_2$ is weakly broken. In contrast, our setup preserves the diagonal subgroup $G_{\rm diag}$. Thus, an open system can be viewed as a particular case of this general scenario in which spurions appear only through combinations preserving $G_{\rm diag}$, specifically $O_1^\dag O_1$, $O_2^\dag O_2$, and $O = O_1 O_2^\dag$ (and its conjugate).\footnote{We do not need to consider additional structures involving derivatives, because they would drop out once the spurions are replaced with their expectation value.} The first two combinations, scalars under $G_1 \times G_2$, modify exactly invariant terms by an amount proportional to the spurion expectation value $\langle O \rangle$. As we will show, these invariant contributions are crucial for ensuring unitarity in the underlying microscopic theory. Consequently, our effective action will depend not only on $O$ but also explicitly on $\langle O \rangle$, slightly departing from the usual spurion formalism. In the Supplemental Material~\cite{supp}]\nocite{Caldeira:1982iu,kamenev2023}, we consider a simple example of a concrete UV completion that motivates our proposal and elucidates the origin of the dependence on $\langle O \rangle$.\footnote{We emphasize that $\langle O \rangle$ is chosen by hand to parametrize the explicit symmetry breaking, and is not an order parameter of the system whose value is to be determined from the dynamics. Its appearance in the EFT is an artifact of the fact that the spurion we introduce does not come in two copies.}.

In the remainder of this letter, we will illustrate our approach in the simplest case of Abelian symmetries, $G = U(1)$. In this case, our spurion field will be a complex field that transforms under $U(1)_1 \times U(1)_2$ like
\begin{align}
	    O \to e^{i (\alpha_1 - \alpha_2)}  O \ ,
\end{align}
ensuring that the spurion expectation value preserves~$U(1)_{\rm diag}$. Gauged $U(1)$ symmetries in open systems---where the off-diagonal symmetry is badly broken---were recently discussed in~\cite{Salcedo:2024nex}.

\section{\texorpdfstring{$U(1)$}{U(1)} Diffusion}

\subsection{Closed systems}

We start by reviewing the effective field theory (EFT) of $U(1)$ diffusion for closed systems in a thermal state. In this case, the Schwinger-Keldysh effective action must be invariant under $U(1)_1 \times U(1)_2$ \emph{spontaneously} broken down to $U(1)_{\rm diag}$~\cite{Akyuz:2023lsm}. The corresponding Goldstone mode, $\varphi_a$, shifts under the ``axial'' $U(1)$, is a singlet under $U(1)_{\rm diag}$, and is  paired with a ``matter field'' $\rho_r$.\footnote{The $a$ and $r$ subscript are introduced to match standard notation in the literature---see e.g.~\cite{Liu:2018kfw}.}.Thermality implies that all correlation functions satisfy the KMS condition. This can be enforced at the level of the effective action by imposing invariance under the dynamical KMS (DKMS) symmetry~\cite{Sieberer:2015hba,Glorioso:2017fpd}:\footnote{The DKMS symmetry relies on a discrete symmetry of the action involving time reversal.  While $CPT$ is a natural candidate for Lorentz-invariant actions, $C$ is spontaneously broken at finite charge density. Thus, implementing DKMS via $PT$, with additional $C$ symmetry imposed through $\varphi_a'(x) = - \varphi_a (x)$ and $\rho_r'(x) = - \rho_r (x)$, becomes more convenient. The transformation rules \eqref{eq: DKMS diffusion} assume invariance under $PT$, and differ from the ones in~\cite{Akyuz:2023lsm} which are built using $CT$.}
\begin{subequations} \label{eq: DKMS diffusion}
\begin{align}
    \varphi_a'(x) &= - \varphi_a(-x) - i \beta \rho_r(-x) + \mathcal{O}(\beta^2), \\
    \rho_r'(x) &=  \rho_r(-x) - \tfrac{i \beta}{4} \partial_t^2 \varphi_a(-x) + \mathcal{O}(\beta^2). \label{eq: DKMS 2}
\end{align}
\end{subequations}
We have expanded these transformation rules in powers of $\beta = 1/T$, anticipating the fact that the effective action will be organized in powers of $E/T$. A discussion of these transformations at all orders in $\beta$ and of the systematics of power counting in $E/T$ can be found in~\cite{Firat:2025upx}.

The Schwinger-Keldysh effective action must also satisfy some constraints that follow from unitarity considerations~\cite{Liu:2018kfw}:
\begin{subequations} \label{eq: unitarity diffusion}
\begin{align}
     & S[\varphi_a = 0, \rho_r] = 0 \ , \\
     & S^*[\varphi_a , \rho_r ] = - S[-\varphi_a , \rho_r ]  \ , \\
     & \Im S \geq 0 \ . \label{eq: unitarity diffusion 3}
\end{align}
\end{subequations}
At lowest order in derivatives and quadratic order in fields, the effective action that satisfies these constraints and is invariant under all the symmetries mentioned above is~\cite{Delacretaz:2023ypv}\footnote{In the incompressible limit, boosts are explicitly broken by neglecting thermal bath phonons. The DKMS symmetry is realized perturbatively up to corrections of $\mathcal{O}(E/T)$.}
\begin{align} \label{eq: closed action}
    S &= \int d^4 x \left[ n(\rho_r) \partial_t \varphi_a + \sigma (\rho_r) \partial_i \varphi_a\left( \frac{i}{\beta}\partial_i\varphi_a  - \partial_i \rho_r   \right) \right] ,
\end{align}
where $d^4 x \equiv dt d^3r$ and $n$ and $\sigma$ are analytic around $\rho_r = 0$, and $\sigma \geqslant 0$ to satisfy \eqref{eq: unitarity diffusion 3}. Imposing invariance under charge conjugation constrains $n$ ($\sigma$) to contain only odd (even) powers of $\rho_r$. Demanding that the terms in \eqref{eq: closed action} are all of the same order implies the power counting relations $\rho_r \sim T \varphi_a$ and $\partial_t \sim \partial_i^2$. Interestingly, the first of these relations also implies that the second term on the right-hand side of \eqref{eq: DKMS 2} is negligible compared to the first one---an approximation usually justified by invoking an expansion in powers of $\hbar$.

The equation of motion obtained by varying $S$ with respect to $\rho_r$ always admits the solution $\varphi_a = 0$, which is the background we will be interested in\footnote{This is more than just a convenient choice, it is also the physical one since it ensures that $\varphi_a$ does not propagate, which would lead to an instability otherwise~\cite{Salcedo:2024nex}. This conclusion can be avoided for purely imaginary saddle points of $\varphi_a$~\cite{kamenev2023}.}. On this background, the equation for $\varphi_a$ reduces to a diffusion equation for $n(\rho_r)$:
\begin{align}
	 \partial_t n (
     \rho_r)= \partial_i \left[ \sigma(\rho_r) \partial^i \rho_r \right] , 
\end{align}
In fact, $n(\rho_r)$ is also the Noether charge density associated with invariance under shifts of $\varphi_a$. These results suggest identifying $\rho_r$ with fluctuations of the chemical potential around equilibrium.

\subsection{Leading corrections in ajar systems}

We will now supplement the action $S$ for a closed system with the leading terms that depend on the spurion field $O$. We will need to ensure that these additional terms are invariant under the DKMS symmetry and satisfy the appropriate unitarity constraints. To this end, we will proceed in two steps. 

First, it will be convenient to think again of our spurion field as being a composite object of the form $O = O_1 O_2^\dag$. This is because the transformation properties of the operators $O_{1,2}$ under the DKMS symmetry are quite simple~\cite{Liu:2018kfw}: 
\begin{align}
	O_{1,2}' (x) = O_{1,2}^\dag (-x) + \mathcal{O} (\beta \partial_t O_{1,2}^\dag) \ ,
\end{align}
where the explicit form of the second term on the righthand side will not be needed because it will vanish when the spurions are replaced with their expectation value. The first two unitarity constraints can also be easily amended to account for the fields $O_{1,2}$~\cite{Liu:2018kfw}:
\begin{subequations}
    \begin{align}
        & S[\varphi_a = 0, \rho_r, O_1, O_2 = O_1] = 0 \ , \\
        & S^*[\varphi_a , \rho_r, O_1, O_2] = - S[-\varphi_a , \rho_r, O_2, O_1 ] \ .
    \end{align}
\end{subequations}
These conditions stem from the unitarity of the microscopic physics describing the combined system and bath, and therefore must be imposed even on the effective action of an open system~\cite{Glorioso:2017fpd}.

Second, following the usual template for coupling Goldstone modes to additional ``matter fields''~\cite{Weinberg:1996kr}, we will find it convenient to work directly with the combination $\tilde O = O e^{- i \varphi_a}$, since it is invariant under  $U(1)_1 \times U(1)_2$. By combining our definitions, we can easily derive the transformation rule for $\tilde O$ and its complex conjugate under DKMS:
\begin{subequations}
    \begin{align}
	\tilde O'(x) &= \tilde O^\dag(-x) + \beta \rho_r (-x) \tilde O^\dag(-x) + \mathcal{O} (\beta \partial_t) \ , \\
        \tilde O^{\dag \prime}(x) &= \tilde O (-x) - \beta \rho_r (-x) \tilde O(-x) + \mathcal{O} (\beta \partial_t) \ ,
\end{align}
\end{subequations}
while the first two unitarity constraints reduce to
\begin{subequations} \label{eq: unitarity constraints diffusion}
    \begin{align}
        & S[\varphi_a = 0, \rho_r, \tilde O \geqslant 0] = 0 \ , \label{eq: unitarity diffusion 1}\\
        & S^*[\varphi_a , \rho_r, \tilde O] = - S[-\varphi_a , \rho_r, \tilde O^\dag ] \ .
    \end{align}
\end{subequations}

At linear order in $O$ and $\langle O \rangle$, there is only one term that is invariant under the DKMS symmetry, and satisfies \emph{all three} unitarity constraints.
\begin{align}
    \Delta S  = \! \frac{i}{\beta} \! \int \! d^4x F(\rho_r) \left[ 2 \langle O \rangle \! - \! \left( \tilde{O}^\dag + \tilde{O} \right)\! + \frac{\beta}{2} \rho_r \left( \tilde{O}^\dag \!- \! \tilde{O} \right) \! \right] ,
\end{align}
where $F(\rho_r)$ is analytic around $\rho_r =0$ and, without loss of generality, such that $F(0) = 1$. Note that the first term proportional to $\langle O \rangle$ is crucial to ensure that the unitarity condition \eqref{eq: unitarity diffusion 1} is satisfied. Replacing $O \to \langle O \rangle = \gamma$, this expression reduces to
\begin{align} \label{eq: leading spurion term diffusion after vev} 
   \!\!\! \Delta S = \int d^4x \, 4 \gamma (\rho_r) \sin \tfrac{\varphi_a}{2}\left( \frac{i}{\beta} \sin \tfrac{\varphi_a}{2} - \frac{\rho_r}{2} \cos \tfrac{\varphi_a}{2} \right) , 
\end{align}
where we have defined $\gamma (\rho_r) \equiv \gamma F(\rho_r)$.

It is easy to check that this expression is invariant under the DKMS transformations \eqref{eq: DKMS diffusion} up to corrections of $\mathcal{O}(E/T)$ and satisfies the unitarity constraints \eqref{eq: unitarity diffusion} provided $\gamma(\rho_r) \geqslant 0$. In the presence of charge conjugation symmetry, $\gamma(\rho_r)$ only contains even powers of $\rho_r$ (under $C$, $O'(x) = O^\dag (x)$).
The nonlinear structure of \eqref{eq: leading spurion term diffusion after vev} implies that, in the limit of weak interaction with the environment, any $n$-point function will introduce only one new free parameter at leading order, i.e. successive measurements of each n-point function will fix the coefficient of the $(n-1)$th parameter in the expansion of $\gamma(\rho_r)$.  We will now present the corrections to the 2- and 3-point functions for the charge density in ajar systems due to the terms in \eqref{eq: leading spurion term diffusion after vev}.

\subsection{Density correlation functions}

The quadratic action for diffusive ajar systems follows from expanding \( S + \Delta S \) to quadratic order in \(\rho_r\) and \(\phi_a \). Inverting these terms yields the 2-point functions for \( \varphi_a \) and \( \rho_r \), or equivalently \( \varphi_a \) and the charge density \( n \):
\begin{align}\label{eq: 2-point functions diffusion}
    \langle n(p) n (-p)\rangle &= \dfrac{2}{\beta}\frac{ \left( \gamma + \sigma k^2 \right)}{  \omega^2 + \left(\gamma/\chi + \mathcal{D} k^2\right)^2 }, \nonumber \\
   \langle n(p) \varphi_a (-p)\rangle &= \frac{1}{  \omega + i\left(\gamma/\chi + \mathcal{D} k^2\right) }, 
\end{align}
where \( p = (\omega, \vec{k}) \), \( \chi \! =  \! dn/d\rho_r|_{\rho_r = 0} \), \( \sigma = \sigma (\rho_r = 0) \), and \( \mathcal{D} \! = \! \sigma / \chi \) is the diffusion coefficient, and delta functions enforcing energy-momentum conservation have been omitted. Notably, environmental interactions gap only the imaginary part of the pole, unlike an approximate \( U(1) \) symmetry, where the real part is also gapped.

The leading 3-point function of charge density arises from expanding \( S + \Delta S \) to cubic order in \( \rho_r \) and \( \varphi_a \). The closed-system result was recently derived in~\cite{Delacretaz:2023ypv}. Expressing interactions in terms of \( n \simeq \chi (\rho_r + \frac{1}{2} \chi' \rho_r^2) \) yields:
\begin{align}
	& S_3 + \Delta S_3  \nonumber \\
    & \quad = \int d^4 x \bigg[ \frac{\sigma \chi'}{\chi^2} n \, \partial_i \varphi_a n  
	+  \sigma' n \, \partial_i \varphi_a  \partial^i\left(\frac{i}{\beta} \varphi_a - \frac{n}{\chi}\right) \nonumber \\
    & \qquad + \frac{\gamma \chi'}{2 \chi^2}  n^2 \varphi_a + \gamma' n \, \varphi_a \left(\frac{i}{\beta} \varphi_a - \frac{n}{\chi}\right) \varphi_a \bigg],  
\end{align}
where primes denote derivatives with respect to \( n \). The resulting three-point function of $n$ is  (with $q_i \equiv \sqrt{\mathcal{D}} k_i$ and $\tilde{\gamma} \equiv \gamma / \chi$):
\begin{widetext}
\begin{align}
    \langle n(p_1) n(p_2) n(p_3) \rangle &=  (\chi T)^2\bigg\{ 12 \tilde{\gamma} \, \frac{\chi'}{\chi} \frac{\left(\tilde{\gamma} + q_1^2\right)\left(\tilde{\gamma} + q_2^2\right)\left(\tilde{\gamma} + q_3^2\right)}{\left[ \omega_1^2 + \left(\tilde{\gamma} +   q_1^2\right)^2 \right]\left[ \omega_2^2 + \left(\tilde{\gamma} +   q_2^2\right)^2 \right]\left[ \omega_3^2 + \left(\tilde{\gamma} +   q_3^2\right)^2 \right]} \nonumber \\
    &\qquad \qquad +\left[ \frac{4\chi'}{\chi} - \frac{2 \sigma'}{\sigma} \right]\frac{\left(\tilde{\gamma} + q_1^2\right)\left(\tilde{\gamma} + q_2^2\right)\left(\tilde{\gamma} + q_3^2\right)( q_1^2 + q_2^2 + q_3^2)}{\left[ \omega_1^2 + \left(\tilde{\gamma} +   q_1^2\right)^2 \right]\left[ \omega_2^2 + \left(\tilde{\gamma} +   q_2^2\right)^2 \right]\left[ \omega_3^2 + \left(\tilde{\gamma} +   q_3^2\right)^2 \right]}  \\
    &\qquad \qquad +\frac{4\sigma'}{ \sigma} \frac{\omega_2 \omega_3(q_2 \cdot q_3) (\tilde{\gamma} + q_1^2)  + \omega_1 \omega_3 (q_1 \cdot q_3) (\tilde{\gamma} + q_2^2) + \omega_1 \omega_2(q_1 \cdot q_2) (\tilde{\gamma} + q_3^2)  }{\left[ \omega_1^2 + \left(\tilde{\gamma} +   q_1^2\right)^2\right]\left[ \omega_2^2 + \left(\tilde{\gamma} +   q_2^2\right)^2\right]\left[ \omega_3^2 + \left(\tilde{\gamma} +   q_3^2\right)^2\right]} \nonumber\\
    &\qquad \qquad - \frac{4\gamma'}{\chi^2} \frac{3 \left[\left(\tilde{\gamma} + q_1^2\right)\left(\tilde{\gamma} + q_2^2\right)\left(\tilde{\gamma} + q_3^2\right)\right] + \omega_2 \omega_3\left(\tilde{\gamma} + q_1^2\right) + \omega_1 \omega_3\left(\tilde{\gamma} + q_2^2\right) + \omega_1 \omega_2\left(\tilde{\gamma} + q_3^2\right)  }{\left[ \omega_1^2 + \left(\tilde{\gamma} +  q_1^2\right)^2\right]\left[ \omega_2^2 + \left(\tilde{\gamma} +   q_2^2\right)^2\right]\left[ \omega_3^2 + \left(\tilde{\gamma} +   q_3^2\right)^2\right]}\bigg\} .   \nonumber
\end{align}
\end{widetext}

\section{\texorpdfstring{$U(1)$}{U(1)} Goldstone Mode}

\subsection{Effective action for ajar systems}

The approach discussed in the previous section applies also when the $U(1)$ symmetry is spontaneously broken. In this case, the corresponding Schwinger-Keldysh effective action for a closed system realizes the entire group $U(1)_1 \times U(1)_2$ nonlinearly. The relevant degrees of freedom are the two Goldstone modes $\pi_r$ and $\pi_a$, which transform under the DKMS symmetry as follows~\cite{Akyuz:2023lsm}:
\begin{subequations} \label{eq: DKMS Goldstone}
\begin{align}
    \pi_a'(x) &= - \pi_a(-x) + i \beta \partial_t \pi_r (-x) + \mathcal{O}(\beta^2), \\
    \pi_r'(x) &=  - \pi_r(-x) - \tfrac{i \beta}{4} \partial_t \pi_a(-x) + \mathcal{O}(\beta^2).
\end{align}
\end{subequations}
The effective action must be invariant under \eqref{eq: DKMS Goldstone} as well as shifts of $\pi_{r,a}$, and must satisfy unitarity constraints of the form \eqref{eq: unitarity diffusion} and \eqref{eq: unitarity constraints diffusion} with $\rho_r \to \partial_t \pi_r, \varphi_a \to \pi_a$. The shift symmetries imply that our fields must enter the effective Lagrangian with at least one derivative. If our system was closed, the effective action would be~\cite{Donos:2023ibv}
\begin{align}
    S &=  \frac{f^2}{c_s^3} \int d^4x \,  \bigg\{ \partial_t \pi_a \partial_t \pi_r - c^2_s \partial_i \pi_a \partial_i \pi_r  \nonumber \\
    & \qquad 
    + \frac{\Sigma_\pi}{f} \, \partial_t \pi_a \left(\frac{i}{\beta}\partial_t \pi_a -\partial_t^2 \pi_r \right)  \\
    & \qquad \quad + \frac{\sigma_\pi c_s^2}{f} \, \partial_i \pi_a \left(\frac{i}{\beta}\partial_i \pi_a -\partial_t \partial_i \pi_r \right) + \cdots \bigg\} , \nonumber
\end{align}
where we have shown explicitly only the terms quadratic in the fields. From the quadratic action,  we deduce the power-counting rules $\pi_a \sim (E/T) \pi_r$ and $\partial_t \sim \partial_i $.

The exchange of charge with an environment explicitly breaks the shift symmetry of $\pi_a$. In ajar systems, where this breaking is soft, it can again be modeled by a spurion $O$, or equivalently $\tilde O = O e^{-i \pi_a}$, that now transforms under DKMS as follows:
\begin{subequations}
    \begin{align}
	\tilde O'(x) &= \tilde O^\dag(-x) - \beta \partial_t \pi_r (-x) \tilde O^\dag(-x) + \mathcal{O} (\beta^2) \ , \\
        \tilde O^{\dag \prime}(x) &= \tilde O (-x) + \beta \partial_t \pi_r (-x) \tilde O(-x) + \mathcal{O} (\beta^2) \ ,
\end{align}
\end{subequations}
At leading order in the spurion field, the effective action receives the following correction
\begin{align}
	\Delta S & =  \frac{i}{\beta} \int d^4 x F(\partial_t \pi_r, \partial_i \pi_r \partial^i \pi_r ) \bigg[ 2 \langle O \rangle - \left( \tilde{O}^\dag + \tilde{O} \right) \nonumber \\
    & \qquad \qquad \qquad \qquad \qquad  + \frac{\beta}{2} \partial_t \pi_r \left( \tilde{O}^\dag - \tilde{O} \right)  \bigg] , \label{eq: spurion action Goldstone}
\end{align}
where, once again, we can assume $F(0,0) = 1$ without loss of generality. 

\subsection{Goldstone correlation functions}

The two-point functions that result from inverting the quadratic part of $S + \Delta S$ are
\begin{align}
    \langle\pi_r \pi_r \rangle&= \frac{2 c_s^3 }{\beta f^2} \frac{\tilde{\gamma} + \frac{1}{f}(\Sigma_\pi \omega^2 + \sigma_\pi c_s^2 k^2)}{\left( \omega^2 \! - \! c_s^2 k^2  \right)^2 \!+ \!\omega^2\left[
    \tilde{\gamma} \! + \! \frac{1}{f}(\Sigma_\pi\omega^2 \!+ \!\sigma_\pi c_s^2 k^2  )\right]^2}, \nonumber\\
    \langle\pi_r  \pi_a \rangle &= \frac{c_s^3}{f^2} \frac{i}{\omega^2 - c_s^2 k^2 + i \tilde{\gamma} \omega  + \frac{i}{f} (\Sigma_\pi \omega^3 \!+ \!  \sigma_\pi c_s^2 \omega k^2)},
\end{align}
where $\tilde{\gamma} \equiv \gamma c_s^3/f^2$. The second correlator is simply the retarded Green's function for the Goldstone mode. 

Once again, we see that ``openness'' adds a finite contribution to the imaginary part of the Goldstone dispersion relation, rather than turning this mode into a pseudo-Goldstone boson by gapping the real part, as would be the case for an approximate symmetry. Physically, it means that the Goldstone mode can only propagate for wave numbers $k \gtrsim \tilde \gamma /c_s$. Higher-point functions for the Goldstone mode can be calculated in a straightforward way by expanding $S+\Delta S$ up to the desired order.

\vspace{.1in}

\begin{acknowledgments}
The authors would like to thank 
Santiago Aguei-Salcedo, Thomas Cole, Luca Delacretaz, Sergei Dubovsky, Eren Firat, Andrew Gomes, Garrett Goon, Filippo Nardi, Alberto Nicolis, Enrico Pajer, Riccardo Rattazzi, and Sergey Sibiryakov for interesting discussions and correspondence. This work is supported by the US Department of Energy Grant No. DE-SC0010118.
\end{acknowledgments}

\putbib

\end{bibunit}

\newpage
\onecolumngrid
\newpage

\section*{Supplementary Material} \label{section:Supp}

\setcounter{equation}{0}
\renewcommand{\theequation}{S.\arabic{equation}}
\renewcommand{\bibnumfmt}[1]{[S#1]}
\renewcommand{\citenumfont}[1]{S#1}

\begin{bibunit}

\subsection*{An Example of a UV Model}

We can illustrate the general approach put forward in this paper by considering a concrete UV completion where a complex scalar field is coupled linearly to a set of complex harmonic oscillators in thermal equilibrium, which we will treat as the bath. This simple system can be thought of as a field theory generalization of the Caldeira-Leggett model~\cite{Caldeira:1982iu}. Our presentation will follow closely that of~\cite{kamenev2023field}. The in-out microscopic Lagrangian is $\mathcal{L} = \mathcal{L}_{sys} + \mathcal{L}_{bath} + \mathcal{L}_{int}$
\begin{subequations} \label{eq: in-out Lagrangian UV model}
\begin{align}
    \mathcal{L}_{sys} & = -\partial_\mu \phi^\dag \partial^\mu \phi - m^2 \phi^\dag \phi  \\
    \mathcal{L}_{bath} &= \sum_s \partial_t \chi_s^\dag \partial_t \chi_s - \Gamma_s^2 \chi_s^\dag \chi_s \\
    \mathcal{L}_{int} &= \sum_s g_s (\chi_s^\dag \phi + \phi^\dag \chi_s).
\end{align}
\end{subequations}
where $\Gamma_s$ and $g_s$ are the frequency and interaction strength of each mode. This action is invariant under a $U(1)$ symmetry acting on $\phi$ and the $\chi_s$'s as follows:
\begin{align} \label{eq: U(1) transformation}
	\phi' = e^{i \alpha} \phi \ , \qquad \qquad \chi_s' = e^{i \alpha} \chi_s \ .
\end{align}

The corresponding microscopic Schwinger-Keldysh action can be obtained by doubling all degrees of freedom and calculating $\mathcal{L}_{\rm SK} = \mathcal{L}(\phi_1, \chi_{s,1}) - \mathcal{L}(\phi_2, \chi_{s,2}) \equiv  \mathcal{L}_{sys}^{\rm SK} + \mathcal{L}_{bath}^{\rm SK} + \mathcal{L}_{int}^{\rm SK}$. Switching from the $(1,2)$ basis to the Schwinger-Keldysh basis of fields,
\begin{align}
	\phi_r = \tfrac{1}{2} \left( \phi_1 + \phi_2 \right) \ , \qquad \phi_a = \phi_1 - \phi_2 \ , \qquad \chi_{s,r} = \tfrac{1}{2} \left( \chi_{s,1} + \chi_{s,2} \right) \ ,  \qquad \chi_{s,a} =  \chi_{s,1} - \chi_{s,2} \ , 
\end{align}
we can express the various components of the Schwinger-Keldysh Lagrangian as follows:
\begin{subequations} \label{eq: L UV model}
\begin{align}
    \mathcal{L}_{sys}^{\rm SK} & = -\partial_\mu \phi_a^\dag \partial^\mu \phi_r - m^2 \phi_a^\dag \phi_r + h.c.  \\
    \mathcal{L}_{bath}^{\rm SK} &= \sum_s \partial_t \chi_{a,s}^\dag \partial_t \chi_{r,s} - \Gamma_s^2 \chi_{a,s}^\dag \chi_{r,s} + h.c.  \\
    \mathcal{L}_{int}^{\rm SK} &= \sum_s g_s \left(\chi_{a,s}^\dag \phi_r + \chi_{r,s}^\dag\phi_a  + h.c.\right).
\end{align}
\end{subequations}
Note that this action is invariant under two copies of the $U(1)$ symmetry, acting separately on the ``1'' and ``2'' fields. Equivalently, the diagonal $U(1)$ symmetry acts simultaneously on the ``$r$'' and ``$a$'' degrees of freedom as in Eq. \eqref{eq: U(1) transformation}, while the second one mixes the ``$r$'' and ``$a$'' fields as follows:
\begin{align}
	\phi_r' = \phi_r \cos \alpha + \tfrac{i}{2} \phi_a \sin \alpha \ , \qquad \qquad \phi_a' = \phi_a \cos \alpha + 2 i \phi_r \sin \alpha \ ,
\end{align}
with a similar action on $\chi_{s, r}$ and  $\chi_{s, a}$, and $\alpha$ the parameter of the $U(1)$ transformation.

Since the Lagrangian \eqref{eq: L UV model} is quadratic, we can integrate out the bath degrees of freedom exactly, treating the $\phi_{r,a}$ fields as sources during the process. This leads to the following (non-local) open effective Lagrangian for the $\phi_{r,a}$'s~\cite{kamenev2023field}:
\begin{align} \label{eq: LSK open}
    \mathcal{L}_{\phi}^{\rm SK} = -\partial_\mu \phi_a^\dag \partial^\mu \phi_r - m^2 \phi_a^\dag \phi_r + h.c.  + \phi_r^\dag D_{ra}^{-1} \phi_a + \phi_a^\dag D_{ar}^{-1} \phi_r +  \phi_a^\dag D_{aa}^{-1} \phi_a 
\end{align}

Let us now assume that the frequencies of the harmonic oscillators have an Ohmic distribution, such that the spectral density can be approximated as follows:
\begin{align}
	J(\omega) =  2 \pi \sum_s (g_s^2 /\Gamma_s)\delta(\omega - \Gamma_s) \overset{\text{Ohmic}}{\longrightarrow} 4 \gamma \omega \ ,
\end{align}
where $\gamma$ is a constant for small frequencies. The Ohmic assumption ensures that the $\phi$'s are able to exchange charge with the bath at arbitrarily small energies, so that, after integrating out the $\chi_s$'s, we are left with an open system. The retarded (advanced) kernels reduce to
\begin{align}
    D^{-1}_{ra(ar)} (\omega)  = \int \frac{d\omega'}{2 \pi} \frac{\omega' J(\omega')}{\omega^{\prime 2} - (\omega \pm i \varepsilon)^2} \quad \overset{\text{Ohmic}}{\longrightarrow}  \quad \text{constant} \pm 2 i \gamma \omega \quad \overset{\text{F.T.}}{\longrightarrow} \quad  \mp 2 \gamma \delta(t-t') \partial_t' \ ,
\end{align}
where the upper (lower) sign applies to the $ra$ ($ar$) kernel. Note that, when carrying out the Fourier transform in the last step we dropped the constant part, since it would simply  renormalize the mass term for $\phi$. In order for $D_{aa}^{-1}$ to become local, we additionally need to take the high temperature limit. Its full non-local form is
\begin{align}
    D^{-1}_{aa} &= \frac{1}{2} \coth\left( \frac{\omega}{2T} \right) \left[ D_{ra}^{-1} - D_{ar}^{-1} \right] \quad \overset{\text{Ohmic}}{\longrightarrow} \quad \left( 4 i \gamma \omega \right) \coth\left( \frac{\omega}{2T} \right) \quad \overset{\beta \rightarrow 0}{\longrightarrow} \quad \left( 2 i \gamma \omega \right)\left( \frac{2T}{\omega} \right) = 4 i T \gamma.
\end{align}
With these approximations, Eq. \eqref{eq: LSK open} reduces to the following Lagrangian that is invariant under the DKMS symmetry at lowest order and describes the open dynamics of our complex scalar field: 
\begin{align} \label{eq: explicit breaking terms 1}
    \mathcal{L}_{open}^{\rm SK} \simeq -\partial_\mu \phi_a^\dag \partial^\mu \phi_r - m^2 \phi_a^\dag \phi_r + h.c. + \frac{2 i\gamma}{\beta} \phi_a^\dag \left(  \phi_a + i \beta \partial_t \phi_r  \right) + \frac{2 i\gamma}{\beta}  \phi_a\left( \phi_a^\dag + i \beta \partial_t \phi_r^\dag  \right)  \ .
\end{align}
Note that this Lagrangian is invariant under the DKMS transformation
\begin{align}
	\phi_r'(x) = \phi_r^\dag(-x) - \tfrac{i \beta}{4} \partial_t \phi_a^\dag (-x) + \mathcal{O} (\beta^2) \ , \qquad \quad \phi_a'(x) = \phi_a^\dag(-x) - i \beta \partial_t \phi_r^\dag (-x) + \mathcal{O} (\beta^2) 
\end{align}
but the last two terms in Eq. \eqref{eq: explicit breaking terms 1} explicitly break $U(1)_1 \times U(1)_2$ down to the diagonal subgroup $U(1)_{\rm diag}$. The strength of this breaking is $\sim\gamma$. It will be helpful to check this by reverting to the $1-2$ basis and focusing on the non-derivative terms. Up to an overall constant, we have
\begin{align} \label{eq: non-derivative symmetry breaking terms}
   \gamma  \phi_a^\dag \phi_a = \gamma (\phi_1^\dag - \phi_2^\dag )(\phi_1 - \phi_2 ) =  \gamma (\phi^\dag_1 \phi_1 + \phi^\dag_2 \phi_2 - \phi_1^\dag \phi_2 - \phi_2^\dag \phi_1 )\ , 
\end{align}
and we see that the last two terms are only invariant under $U(1)_{\rm diag}$.

Following the spurion logic, we can make the above term invariant by replacing the symmetry breaking parameter $\gamma$ with two complex spurion fields $O_1$ and $O_2$ that couple to the $\phi_i$'s as follows:
\begin{align}  \label{eq: non-derivative symmetry breaking terms with spurions}
     O_1^\dag O_1 \phi_1^\dag \phi_1 +  O_2^\dag O_2 \phi_2^\dag \phi_2 - O^\dag_2 O_1 \phi_1^\dag \phi_2 - O^\dag_1 O_2 \phi_2^\dag \phi_1 \ .
\end{align}
To recover the terms in \eqref{eq: non-derivative symmetry breaking terms}, the spurion vev should be $\langle O_1 \rangle = \langle O_2 \rangle = \sqrt{\gamma}$. Note that the spurion fields appear only in the quadratic combinations discussed in the Introduction---there are no linear terms in $O_1$ or $O_2$.  

Adding a symmetry breaking potential for $\phi$ to the original Lagrangian \eqref{eq: in-out Lagrangian UV model} would not modify the argument that leads to \eqref{eq: non-derivative symmetry breaking terms with spurions}, since the $\phi_i$'s were treated as external sources while integrating the bath degrees of freedom. In this case, we can parametrize $\phi_i = (v + \sigma_i) e^{i \pi_i}$, where $v$ is the expectation value of the complex scalar. After integrating out the radial modes  $\sigma_i$ at tree level, the expression \eqref{eq: non-derivative symmetry breaking terms with spurions} reduces to
\begin{align}
   v^2 \left( O_1^\dag O_1 +  O_2^\dag O_2 -  O^\dag_2 O_1 e^{-i  \pi_a} - O^\dag_1 O_2  e^{ i \pi_a}\right) , 
\end{align}
where we have introduced $\pi_a = \pi_1 - \pi_2$. A moment of reflection shows that this expression coincides with the combination of non-derivative terms in the square bracket of \eqref{eq: spurion action Goldstone}.

\putbib
    
\end{bibunit}

\begin{thebibliography}{0}%
\makeatletter
\providecommand \@ifxundefined [1]{%
 \@ifx{#1\undefined}
}%
\providecommand \@ifnum [1]{%
 \ifnum #1\expandafter \@firstoftwo
 \else \expandafter \@secondoftwo
 \fi
}%
\providecommand \@ifx [1]{%
 \ifx #1\expandafter \@firstoftwo
 \else \expandafter \@secondoftwo
 \fi
}%
\providecommand \natexlab [1]{#1}%
\providecommand \enquote  [1]{``#1''}%
\providecommand \bibnamefont  [1]{#1}%
\providecommand \bibfnamefont [1]{#1}%
\providecommand \citenamefont [1]{#1}%
\providecommand \href@noop [0]{\@secondoftwo}%
\providecommand \href [0]{\begingroup \@sanitize@url \@href}%
\providecommand \@href[1]{\@@startlink{#1}\@@href}%
\providecommand \@@href[1]{\endgroup#1\@@endlink}%
\providecommand \@sanitize@url [0]{\catcode `\\12\catcode `\$12\catcode `\&12\catcode `\#12\catcode `\^12\catcode `\_12\catcode `\%12\relax}%
\providecommand \@@startlink[1]{}%
\providecommand \@@endlink[0]{}%
\providecommand \url  [0]{\begingroup\@sanitize@url \@url }%
\providecommand \@url [1]{\endgroup\@href {#1}{\urlprefix }}%
\providecommand \urlprefix  [0]{URL }%
\providecommand \Eprint [0]{\href }%
\providecommand \doibase [0]{https://doi.org/}%
\providecommand \selectlanguage [0]{\@gobble}%
\providecommand \bibinfo  [0]{\@secondoftwo}%
\providecommand \bibfield  [0]{\@secondoftwo}%
\providecommand \translation [1]{[#1]}%
\providecommand \BibitemOpen [0]{}%
\providecommand \bibitemStop [0]{}%
\providecommand \bibitemNoStop [0]{.\EOS\space}%
\providecommand \EOS [0]{\spacefactor3000\relax}%
\providecommand \BibitemShut  [1]{\csname bibitem#1\endcsname}%
\let\auto@bib@innerbib\@empty
\end{thebibliography}%


\providecommand{\noopsort}[1]{}\providecommand{\singleletter}[1]{#1}%
\begin{thebibliography}{20}%
\makeatletter
\providecommand \@ifxundefined [1]{%
 \@ifx{#1\undefined}
}%
\providecommand \@ifnum [1]{%
 \ifnum #1\expandafter \@firstoftwo
 \else \expandafter \@secondoftwo
 \fi
}%
\providecommand \@ifx [1]{%
 \ifx #1\expandafter \@firstoftwo
 \else \expandafter \@secondoftwo
 \fi
}%
\providecommand \natexlab [1]{#1}%
\providecommand \enquote  [1]{``#1''}%
\providecommand \bibnamefont  [1]{#1}%
\providecommand \bibfnamefont [1]{#1}%
\providecommand \citenamefont [1]{#1}%
\providecommand \href@noop [0]{\@secondoftwo}%
\providecommand \href [0]{\begingroup \@sanitize@url \@href}%
\providecommand \@href[1]{\@@startlink{#1}\@@href}%
\providecommand \@@href[1]{\endgroup#1\@@endlink}%
\providecommand \@sanitize@url [0]{\catcode `\\12\catcode `\$12\catcode `\&12\catcode `\#12\catcode `\^12\catcode `\_12\catcode `\%12\relax}%
\providecommand \@@startlink[1]{}%
\providecommand \@@endlink[0]{}%
\providecommand \url  [0]{\begingroup\@sanitize@url \@url }%
\providecommand \@url [1]{\endgroup\@href {#1}{\urlprefix }}%
\providecommand \urlprefix  [0]{URL }%
\providecommand \Eprint [0]{\href }%
\providecommand \doibase [0]{https://doi.org/}%
\providecommand \selectlanguage [0]{\@gobble}%
\providecommand \bibinfo  [0]{\@secondoftwo}%
\providecommand \bibfield  [0]{\@secondoftwo}%
\providecommand \translation [1]{[#1]}%
\providecommand \BibitemOpen [0]{}%
\providecommand \bibitemStop [0]{}%
\providecommand \bibitemNoStop [0]{.\EOS\space}%
\providecommand \EOS [0]{\spacefactor3000\relax}%
\providecommand \BibitemShut  [1]{\csname bibitem#1\endcsname}%
\let\auto@bib@innerbib\@empty
\bibitem [{\citenamefont {Haehl}\ \emph {et~al.}(2017)\citenamefont {Haehl}, \citenamefont {Loganayagam},\ and\ \citenamefont {Rangamani}}]{Haehl:2016pec}%
  \BibitemOpen
  \bibfield  {author} {\bibinfo {author} {\bibfnamefont {F.~M.}\ \bibnamefont {Haehl}}, \bibinfo {author} {\bibfnamefont {R.}~\bibnamefont {Loganayagam}},\ and\ \bibinfo {author} {\bibfnamefont {M.}~\bibnamefont {Rangamani}},\ }\href@noop {} {\bibfield  {journal} {\bibinfo  {journal} {JHEP}\ }\textbf {\bibinfo {volume} {06}},\ \bibinfo {pages} {069}},\ \Eprint {https://arxiv.org/abs/1610.01940} {arXiv:1610.01940 [hep-th]} \BibitemShut {NoStop}%
\bibitem [{\citenamefont {Liu}\ and\ \citenamefont {Glorioso}(2018)}]{Liu:2018kfw}%
  \BibitemOpen
  \bibfield  {author} {\bibinfo {author} {\bibfnamefont {H.}~\bibnamefont {Liu}}\ and\ \bibinfo {author} {\bibfnamefont {P.}~\bibnamefont {Glorioso}},\ }\href {https://doi.org/10.22323/1.305.0008} {\bibfield  {journal} {\bibinfo  {journal} {PoS}\ }\textbf {\bibinfo {volume} {TASI2017}},\ \bibinfo {pages} {008} (\bibinfo {year} {2018})},\ \Eprint {https://arxiv.org/abs/1805.09331} {arXiv:1805.09331 [hep-th]} \BibitemShut {NoStop}%
\bibitem [{\citenamefont {Salcedo}\ \emph {et~al.}(2024)\citenamefont {Salcedo}, \citenamefont {Colas},\ and\ \citenamefont {Pajer}}]{Salcedo:2024smn}%
  \BibitemOpen
  \bibfield  {author} {\bibinfo {author} {\bibfnamefont {S.~A.}\ \bibnamefont {Salcedo}}, \bibinfo {author} {\bibfnamefont {T.}~\bibnamefont {Colas}},\ and\ \bibinfo {author} {\bibfnamefont {E.}~\bibnamefont {Pajer}},\ }\href {https://doi.org/10.1007/JHEP10(2024)248} {\bibfield  {journal} {\bibinfo  {journal} {JHEP}\ }\textbf {\bibinfo {volume} {10}},\ \bibinfo {pages} {248}},\ \Eprint {https://arxiv.org/abs/2404.15416} {arXiv:2404.15416 [hep-th]} \BibitemShut {NoStop}%
\bibitem [{\citenamefont {Kamenev}(2023)}]{kamenev2023}%
  \BibitemOpen
  \bibfield  {author} {\bibinfo {author} {\bibfnamefont {A.}~\bibnamefont {Kamenev}},\ }\href@noop {} {\emph {\bibinfo {title} {Field theory of non-equilibrium systems, {\it 2nd ed.}}}}\ (\bibinfo  {publisher} {Cambridge University Press},\ \bibinfo {year} {2023})\BibitemShut {NoStop}%
\bibitem [{\citenamefont {Haehl}\ and\ \citenamefont {Rangamani}(2024)}]{Haehl:2024pqu}%
  \BibitemOpen
  \bibfield  {author} {\bibinfo {author} {\bibfnamefont {F.~M.}\ \bibnamefont {Haehl}}\ and\ \bibinfo {author} {\bibfnamefont {M.}~\bibnamefont {Rangamani}}\ }(\bibinfo {year} {2024})\ \Eprint {https://arxiv.org/abs/2410.10602} {arXiv:2410.10602 [hep-th]} \BibitemShut {NoStop}%
\bibitem [{\citenamefont {Sieberer}\ \emph {et~al.}(2016)\citenamefont {Sieberer}, \citenamefont {Buchhold},\ and\ \citenamefont {Diehl}}]{Sieberer:2015svu}%
  \BibitemOpen
  \bibfield  {author} {\bibinfo {author} {\bibfnamefont {L.~M.}\ \bibnamefont {Sieberer}}, \bibinfo {author} {\bibfnamefont {M.}~\bibnamefont {Buchhold}},\ and\ \bibinfo {author} {\bibfnamefont {S.}~\bibnamefont {Diehl}},\ }\href {https://doi.org/10.1088/0034-4885/79/9/096001} {\bibfield  {journal} {\bibinfo  {journal} {Rept. Prog. Phys.}\ }\textbf {\bibinfo {volume} {79}},\ \bibinfo {pages} {096001} (\bibinfo {year} {2016})},\ \Eprint {https://arxiv.org/abs/1512.00637} {arXiv:1512.00637 [cond-mat.quant-gas]} \BibitemShut {NoStop}%
\bibitem [{\citenamefont {Akyuz}\ \emph {et~al.}(2024)\citenamefont {Akyuz}, \citenamefont {Goon},\ and\ \citenamefont {Penco}}]{Akyuz:2023lsm}%
  \BibitemOpen
  \bibfield  {author} {\bibinfo {author} {\bibfnamefont {C.~O.}\ \bibnamefont {Akyuz}}, \bibinfo {author} {\bibfnamefont {G.}~\bibnamefont {Goon}},\ and\ \bibinfo {author} {\bibfnamefont {R.}~\bibnamefont {Penco}},\ }\href {https://doi.org/10.1007/JHEP06(2024)004} {\bibfield  {journal} {\bibinfo  {journal} {JHEP}\ }\textbf {\bibinfo {volume} {06}},\ \bibinfo {pages} {004}},\ \Eprint {https://arxiv.org/abs/2306.17232} {arXiv:2306.17232 [hep-th]} \BibitemShut {NoStop}%
\bibitem [{\citenamefont {Hongo}\ \emph {et~al.}(2021)\citenamefont {Hongo}, \citenamefont {Kim}, \citenamefont {Noumi},\ and\ \citenamefont {Ota}}]{Hongo:2019qhi}%
  \BibitemOpen
  \bibfield  {author} {\bibinfo {author} {\bibfnamefont {M.}~\bibnamefont {Hongo}}, \bibinfo {author} {\bibfnamefont {S.}~\bibnamefont {Kim}}, \bibinfo {author} {\bibfnamefont {T.}~\bibnamefont {Noumi}},\ and\ \bibinfo {author} {\bibfnamefont {A.}~\bibnamefont {Ota}},\ }\href {https://doi.org/10.1103/PhysRevD.103.056020} {\bibfield  {journal} {\bibinfo  {journal} {Phys. Rev. D}\ }\textbf {\bibinfo {volume} {103}},\ \bibinfo {pages} {056020} (\bibinfo {year} {2021})},\ \Eprint {https://arxiv.org/abs/1907.08609} {arXiv:1907.08609 [hep-th]} \BibitemShut {NoStop}%
\bibitem [{\citenamefont {Penco}(2020)}]{Penco:2020kvy}%
  \BibitemOpen
  \bibfield  {author} {\bibinfo {author} {\bibfnamefont {R.}~\bibnamefont {Penco}},\ }\href@noop {} {\  (\bibinfo {year} {2020})},\ \Eprint {https://arxiv.org/abs/2006.16285} {arXiv:2006.16285 [hep-th]} \BibitemShut {NoStop}%
\bibitem [{\citenamefont {Hongo}\ \emph {et~al.}(2024)\citenamefont {Hongo}, \citenamefont {Sogabe}, \citenamefont {Stephanov},\ and\ \citenamefont {Yee}}]{Hongo:2024brb}%
  \BibitemOpen
  \bibfield  {author} {\bibinfo {author} {\bibfnamefont {M.}~\bibnamefont {Hongo}}, \bibinfo {author} {\bibfnamefont {N.}~\bibnamefont {Sogabe}}, \bibinfo {author} {\bibfnamefont {M.~A.}\ \bibnamefont {Stephanov}},\ and\ \bibinfo {author} {\bibfnamefont {H.-U.}\ \bibnamefont {Yee}},\ }\href@noop {} {\  (\bibinfo {year} {2024})},\ \Eprint {https://arxiv.org/abs/2411.08016} {arXiv:2411.08016 [hep-th]} \BibitemShut {NoStop}%
\bibitem [{\citenamefont {Baggioli}\ \emph {et~al.}(2023)\citenamefont {Baggioli}, \citenamefont {Bu},\ and\ \citenamefont {Ziogas}}]{Baggioli:2023tlc}%
  \BibitemOpen
  \bibfield  {author} {\bibinfo {author} {\bibfnamefont {M.}~\bibnamefont {Baggioli}}, \bibinfo {author} {\bibfnamefont {Y.}~\bibnamefont {Bu}},\ and\ \bibinfo {author} {\bibfnamefont {V.}~\bibnamefont {Ziogas}},\ }\href {https://doi.org/10.1007/JHEP09(2023)019} {\bibfield  {journal} {\bibinfo  {journal} {JHEP}\ }\textbf {\bibinfo {volume} {09}},\ \bibinfo {pages} {019}},\ \Eprint {https://arxiv.org/abs/2304.14173} {arXiv:2304.14173 [hep-th]} \BibitemShut {NoStop}%
\bibitem [{sup()}]{supp}%
  \BibitemOpen
  \href@noop {} {}\bibinfo {note} {See Supplemental Material at http://link.aps.org/supplemental/10.1103/mrzp-zh4v, which includes Ref. [13], for an example of a UV model that motivates our spurion construction}\BibitemShut {NoStop}%
\bibitem [{\citenamefont {Caldeira}\ and\ \citenamefont {Leggett}(1983)}]{Caldeira:1982iu}%
  \BibitemOpen
  \bibfield  {author} {\bibinfo {author} {\bibfnamefont {A.~O.}\ \bibnamefont {Caldeira}}\ and\ \bibinfo {author} {\bibfnamefont {A.~J.}\ \bibnamefont {Leggett}},\ }\href@noop {} {\bibfield  {journal} {\bibinfo  {journal} {Physica}\ }\textbf {\bibinfo {volume} {121A}},\ \bibinfo {pages} {587} (\bibinfo {year} {1983})}\BibitemShut {NoStop}%
\bibitem [{\citenamefont {Salcedo}\ \emph {et~al.}(2025)\citenamefont {Salcedo}, \citenamefont {Colas},\ and\ \citenamefont {Pajer}}]{Salcedo:2024nex}%
  \BibitemOpen
  \bibfield  {author} {\bibinfo {author} {\bibfnamefont {S.~A.}\ \bibnamefont {Salcedo}}, \bibinfo {author} {\bibfnamefont {T.}~\bibnamefont {Colas}},\ and\ \bibinfo {author} {\bibfnamefont {E.}~\bibnamefont {Pajer}},\ }\href {https://doi.org/10.1007/JHEP03(2025)138} {\bibfield  {journal} {\bibinfo  {journal} {JHEP}\ }\textbf {\bibinfo {volume} {03}},\ \bibinfo {pages} {138}},\ \Eprint {https://arxiv.org/abs/2412.12299} {arXiv:2412.12299 [hep-th]} \BibitemShut {NoStop}%
\bibitem [{\citenamefont {Sieberer}\ \emph {et~al.}(2015)\citenamefont {Sieberer}, \citenamefont {Chiocchetta}, \citenamefont {Gambassi}, \citenamefont {T\"auber},\ and\ \citenamefont {Diehl}}]{Sieberer:2015hba}%
  \BibitemOpen
  \bibfield  {author} {\bibinfo {author} {\bibfnamefont {L.~M.}\ \bibnamefont {Sieberer}}, \bibinfo {author} {\bibfnamefont {A.}~\bibnamefont {Chiocchetta}}, \bibinfo {author} {\bibfnamefont {A.}~\bibnamefont {Gambassi}}, \bibinfo {author} {\bibfnamefont {U.~C.}\ \bibnamefont {T\"auber}},\ and\ \bibinfo {author} {\bibfnamefont {S.}~\bibnamefont {Diehl}},\ }\href@noop {} {\bibfield  {journal} {\bibinfo  {journal} {Phys. Rev. B}\ }\textbf {\bibinfo {volume} {92}},\ \bibinfo {pages} {134307} (\bibinfo {year} {2015})},\ \Eprint {https://arxiv.org/abs/1505.00912} {arXiv:1505.00912 [cond-mat.stat-mech]} \BibitemShut {NoStop}%
\bibitem [{\citenamefont {Glorioso}\ \emph {et~al.}(2017)\citenamefont {Glorioso}, \citenamefont {Crossley},\ and\ \citenamefont {Liu}}]{Glorioso:2017fpd}%
  \BibitemOpen
  \bibfield  {author} {\bibinfo {author} {\bibfnamefont {P.}~\bibnamefont {Glorioso}}, \bibinfo {author} {\bibfnamefont {M.}~\bibnamefont {Crossley}},\ and\ \bibinfo {author} {\bibfnamefont {H.}~\bibnamefont {Liu}},\ }\href@noop {} {\bibfield  {journal} {\bibinfo  {journal} {JHEP}\ }\textbf {\bibinfo {volume} {09}},\ \bibinfo {pages} {096}},\ \Eprint {https://arxiv.org/abs/1701.07817} {arXiv:1701.07817 [hep-th]} \BibitemShut {NoStop}%
\bibitem [{\citenamefont {Firat}\ \emph {et~al.}(2025)\citenamefont {Firat}, \citenamefont {Gomes}, \citenamefont {Nardi}, \citenamefont {Penco},\ and\ \citenamefont {Rattazzi}}]{Firat:2025upx}%
  \BibitemOpen
  \bibfield  {author} {\bibinfo {author} {\bibfnamefont {E.}~\bibnamefont {Firat}}, \bibinfo {author} {\bibfnamefont {A.}~\bibnamefont {Gomes}}, \bibinfo {author} {\bibfnamefont {F.}~\bibnamefont {Nardi}}, \bibinfo {author} {\bibfnamefont {R.}~\bibnamefont {Penco}},\ and\ \bibinfo {author} {\bibfnamefont {R.}~\bibnamefont {Rattazzi}},\ }\href@noop {} {\  (\bibinfo {year} {2025})},\ \Eprint {https://arxiv.org/abs/2508.18346} {arXiv:2508.18346 [hep-th]} \BibitemShut {NoStop}%
\bibitem [{\citenamefont {Delacretaz}\ and\ \citenamefont {Mishra}(2024)}]{Delacretaz:2023ypv}%
  \BibitemOpen
  \bibfield  {author} {\bibinfo {author} {\bibfnamefont {L.~V.}\ \bibnamefont {Delacretaz}}\ and\ \bibinfo {author} {\bibfnamefont {R.}~\bibnamefont {Mishra}},\ }\href {https://doi.org/10.21468/SciPostPhys.16.2.047} {\bibfield  {journal} {\bibinfo  {journal} {SciPost Phys.}\ }\textbf {\bibinfo {volume} {16}},\ \bibinfo {pages} {047} (\bibinfo {year} {2024})},\ \Eprint {https://arxiv.org/abs/2304.03236} {arXiv:2304.03236 [cond-mat.str-el]} \BibitemShut {NoStop}%
\bibitem [{\citenamefont {Weinberg}(2013)}]{Weinberg:1996kr}%
  \BibitemOpen
  \bibfield  {author} {\bibinfo {author} {\bibfnamefont {S.}~\bibnamefont {Weinberg}},\ }\href@noop {} {\emph {\bibinfo {title} {{The quantum theory of fields. Vol. 2: Modern applications}}}}\ (\bibinfo  {publisher} {Cambridge University Press},\ \bibinfo {year} {2013})\BibitemShut {NoStop}%
\bibitem [{\citenamefont {Donos}\ and\ \citenamefont {Kailidis}(2024)}]{Donos:2023ibv}%
  \BibitemOpen
  \bibfield  {author} {\bibinfo {author} {\bibfnamefont {A.}~\bibnamefont {Donos}}\ and\ \bibinfo {author} {\bibfnamefont {P.}~\bibnamefont {Kailidis}},\ }\href {https://doi.org/10.1007/JHEP01(2024)110} {\bibfield  {journal} {\bibinfo  {journal} {JHEP}\ }\textbf {\bibinfo {volume} {01}},\ \bibinfo {pages} {110}},\ \Eprint {https://arxiv.org/abs/2304.06008} {arXiv:2304.06008 [hep-th]} \BibitemShut {NoStop}%
\end{thebibliography}%


\providecommand{\noopsort}[1]{}\providecommand{\singleletter}[1]{#1}%
\begin{thebibliography}{2}%
\makeatletter
\providecommand \@ifxundefined [1]{%
 \@ifx{#1\undefined}
}%
\providecommand \@ifnum [1]{%
 \ifnum #1\expandafter \@firstoftwo
 \else \expandafter \@secondoftwo
 \fi
}%
\providecommand \@ifx [1]{%
 \ifx #1\expandafter \@firstoftwo
 \else \expandafter \@secondoftwo
 \fi
}%
\providecommand \natexlab [1]{#1}%
\providecommand \enquote  [1]{``#1''}%
\providecommand \bibnamefont  [1]{#1}%
\providecommand \bibfnamefont [1]{#1}%
\providecommand \citenamefont [1]{#1}%
\providecommand \href@noop [0]{\@secondoftwo}%
\providecommand \href [0]{\begingroup \@sanitize@url \@href}%
\providecommand \@href[1]{\@@startlink{#1}\@@href}%
\providecommand \@@href[1]{\endgroup#1\@@endlink}%
\providecommand \@sanitize@url [0]{\catcode `\\12\catcode `\$12\catcode `\&12\catcode `\#12\catcode `\^12\catcode `\_12\catcode `\%12\relax}%
\providecommand \@@startlink[1]{}%
\providecommand \@@endlink[0]{}%
\providecommand \url  [0]{\begingroup\@sanitize@url \@url }%
\providecommand \@url [1]{\endgroup\@href {#1}{\urlprefix }}%
\providecommand \urlprefix  [0]{URL }%
\providecommand \Eprint [0]{\href }%
\providecommand \doibase [0]{https://doi.org/}%
\providecommand \selectlanguage [0]{\@gobble}%
\providecommand \bibinfo  [0]{\@secondoftwo}%
\providecommand \bibfield  [0]{\@secondoftwo}%
\providecommand \translation [1]{[#1]}%
\providecommand \BibitemOpen [0]{}%
\providecommand \bibitemStop [0]{}%
\providecommand \bibitemNoStop [0]{.\EOS\space}%
\providecommand \EOS [0]{\spacefactor3000\relax}%
\providecommand \BibitemShut  [1]{\csname bibitem#1\endcsname}%
\let\auto@bib@innerbib\@empty
\bibitem [{\citenamefont {Caldeira}\ and\ \citenamefont {Leggett}(1983)}]{Caldeira:1982iu}%
  \BibitemOpen
  \bibfield  {author} {\bibinfo {author} {\bibfnamefont {A.~O.}\ \bibnamefont {Caldeira}}\ and\ \bibinfo {author} {\bibfnamefont {A.~J.}\ \bibnamefont {Leggett}},\ }\href@noop {} {\bibfield  {journal} {\bibinfo  {journal} {Physica}\ }\textbf {\bibinfo {volume} {121A}},\ \bibinfo {pages} {587} (\bibinfo {year} {1983})}\BibitemShut {NoStop}%
\bibitem [{\citenamefont {Kamenev}(2023)}]{kamenev2023field}%
  \BibitemOpen
  \bibfield  {author} {\bibinfo {author} {\bibfnamefont {A.}~\bibnamefont {Kamenev}},\ }\href@noop {} {\emph {\bibinfo {title} {Field Theory of Non-Equilibrium Systems}}},\ \bibinfo {edition} {2nd}\ ed.\ (\bibinfo  {publisher} {Cambridge University Press},\ \bibinfo {year} {2023})\BibitemShut {NoStop}%
\end{thebibliography}%
\end{document}